\documentclass{aa}
\usepackage{graphicx}
\usepackage{latexsym}
\usepackage{amssymb} 
\usepackage[authoryear,sort]{natbib}
\bibpunct{(}{)}{;}{a}{}{,}
\usepackage{txfonts}
\def\etal{et al.}

\def\teff{\ifmmode T_{\rm eff} \else $T_{\mathrm{eff}}$\fi}

\def\ltsima{$\buildrel<\over\sim$}
\def\lsim{\lower.5ex\hbox{\ltsima}}
\newcommand{\hi}{H~{\sc i}}
\newcommand{\hii}{H~{\sc ii}}
\newcommand{\ha}{\ifmmode {\rm H}\alpha \else H$\alpha$\fi}
\newcommand{\hb}{\ifmmode {\rm H}\beta \else H$\beta$\fi}
\newcommand{\lya}{\ifmmode {\rm Ly}\alpha \else Ly$\alpha$\fi}

\def\kms{km s$^{-1}$}

\def\msun{\ifmmode M_{\odot} \else M$_{\odot}$\fi}
\def\msunyr{\ifmmode M_{\odot} {\rm yr}^{-1} \else M$_{\odot}$ yr$^{-1}$\fi}
\def\zsun{\ifmmode Z_{\odot} \else Z$_{\odot}$\fi}
\def\lsun{\ifmmode L_{\odot} \else L$_{\odot}$\fi}

\def\mup{\ifmmode M_{\rm up} \else M$_{\rm up}$\fi}
\def\mlow{\ifmmode M_{\rm low} \else M$_{\rm low}$\fi}

\newcommand{\oh}{\ifmmode 12 + \log({\rm O/H}) \else$12 + \log({\rm
O/H})$\fi}
\def\flyf{\ifmmode f_{\rm Lyf} \else $f_{\rm Lyf}$\fi}
\def\pz{\ifmmode P(z) \else $P(z)$\fi}
\def\ki2{\ifmmode \chi^2 \else $\chi^2$\fi}
\def\zphot{\ifmmode z_{\rm phot} \else $z_{\rm phot}$\fi}

\newcommand{\xphot}{\ifmmode x_\gamma \else $v_\gamma$\fi}
\newcommand{\xobs}{\ifmmode x_{\rm obs} \else $x_{\rm obs}$\fi}
\newcommand{\xcmf}{\ifmmode x_{\rm CMF} \else $x_{\rm CMF}$\fi}
\newcommand{\vexp}{\ifmmode V_{\rm exp} \else $V_{\rm exp}$\fi}
\newcommand{\vmax}{\ifmmode V_{\rm max} \else $V_{\rm max}$\fi}
\newcommand{\vback}{\ifmmode V_{\rm back} \else $V_{\rm back}$\fi}
\newcommand{\nh}{\ifmmode N_{\rm H} \else $N_{\rm H}$\fi}

\begin{document}
\title{3D \lya\ radiation transfer. II. 
  Fitting the Lyman break galaxy MS 1512--cB58 
	and implications for \lya\ emission in high-$z$ starbursts}

\author{Daniel Schaerer\inst{1,2}, Anne Verhamme\inst{1}}
\offprints{daniel.schaerer@obs.unige.ch}
\titlerunning{3D \lya\ radiation transfer. II. cB58 and implications for \lya\
in high-$z$ starbursts}

\institute{
Observatoire de Gen\`eve, Universit\'e de Gen\`eve,
51, Ch. des Maillettes, 
CH--1290 Sauverny, Switzerland
\and
Laboratoire d'Astrophysique (UMR 5572),
Observatoire Midi-Pyr\'en\'ees,
14 Avenue E. Belin,
F--31400 Toulouse, France}
\date{Received date; accepted date}

\abstract{}{
To understand the origin of the \lya\ line profile of the prototypical
Lyman break galaxy (LBG) MS 1512--cB58 and other similar objects. To attempt
a consistent fit of \lya\ and other UV properties of this galaxy.
To understand the main physical parameter(s) responsible for the large variation 
of \lya\ line strengths and profiles observed in LBGs.}
{
Using our 3D \lya\ radiation transfer code (Verhamme \etal\ 2006), we
compute the radiation transfer of \lya\ and UV continuum photons
including dust. Observational constraints on the neutral gas
(column density, kinematics, etc.) are taken from other analysis
of this object.}
{
The observed \lya\ profile of MS 1512--cB58 is reproduced for the first time
taking radiation transfer and all observational constraints into account.
The observed absorption profile is found to result naturally
from the observed amount of dust and the relatively high HI column density \nh.
Radiation transfer effects and suppresion by dust transform a
strong intrinsic \lya\ emission with EW(\lya) $\ga$ 60 \AA\ into the
observed faint superposed \lya\ emission peak.
We propose that the vast majority of LBGs have intrinsically
EW(\lya) $\sim$ 60--80 \AA\ or larger, and that the main physical parameter
responsible for the observed variety of \lya\ strengths and profiles
in LBGs is \nh\ and the accompanying variation of the dust content.
Observed EW(\lya) distributions, \lya\ luminosity functions,
and related quantities must therefore be corrected for radiation transfer
and dust effects. The implications from our scenario on the duty-cycle
of \lya\ emitters are also discussed.
}{}


\maketitle
\section{Introduction}
\label{s_intro}
Discovered serendipitously by \citet{Yee96}, the Lyman
Break Galaxy (LBG) MS 1512--cB58 (hereafter cB58) located at 
$z\sim 2.73$ is strongly gravitationally magnified by the foreground
cluster MS 1512+36. It is by far the best studied LBG, which
has been observed at many wavelengths
\citep[see][]{Elli96,Fray97,Bech97,Naka97,Pett00,Tepl00,Bake01,Sawi01,
Sava02,Pett02,Bake04,Tepl04}. In particular, rest-frame UV 
\citep{Pett00,Pett02}, optical \citep{Tepl00,Tepl04}, and radio
\citep{Bake01, Bake04} observations have allowed to draw a fairly
global picture of this star-forming galaxy and its interstellar medium 
(ISM): it is young ($\sim$ 50--300 Myr), 
massive ($\sim 10^{10}$ \msun), with a sub-solar metallicity ($Z \sim 2/5$ \zsun),
a star formation rate (SFR $\sim 24$ \msunyr) typical of LBGs, and
it is surrounded by outflowing material at $\vexp \sim 250$ \kms. 

Thanks to its apparent brightness the rest-UV spectra of this galaxy
are of comparable quality as the best studied local starbursts.
This has in particular made it possible to detect numerous 
stellar and interstellar (IS) lines, to study their kinematics,
and to use them for detailed UV line fits to constrain its stellar
content, age, star formation history etc. \citep[e.g.][]{Elli96,Pett00,deMell00}.
Apart from its brightness, magnified by a factor $\sim$ 30 thanks to
gravitational lensing \citep{Seitz98},  the properties of cB58 are 
empirically found to be typical of LBGs \citep{Shapley03}. 
More precisely, in the sample of \citet{Shapley03} cB58 is found in 
the quartile ($\sim 25$ \%) of LBGs with the strongest \lya\ absorption.
As such a representative and given the quality and amout of observational
data available, cB58 is of particular interest.

Our objectives are to examine if the observed 
\lya\ line profile and strength of cB58 can be understood, and if one is able to
obtain consistent diagnostics from the \lya\ line and from the other UV
and broad band features used in earlier papers.
Such an attempt is carried out here for the first time with a \lya\ radiation transfer
code \citep{Verh06}. Other LBGs with varying strengths of \lya\ emission 
will be analysed in a subsequent paper \citep{Verh07}.
In particular we wish to elucidate what causes the broad \lya\
absorption and very weak superposed \lya\ emission in cB58
and in other LBGs showing similar \lya\ spectra \citep[cf.][]{Shapley03}.
Similarly we want to understand what physical parameter(s) distinguish 
the different subtypes of LBGs showing large variations 
especially in \lya.
Finally, fitting observed starburst spectra with detailed \lya\ radiation 
transfer modeling should in general provide insight for the
use of \lya\ as a diagnostic in distant galaxies.

The remainder of the paper is structured as follows.
The main observational properties of cB58 are summarised in Sect.\ \ref{s_obs}.
The principles of our simulations and fit method are described
in Sect.\ \ref{s_fit}.
In Sect.\ \ref{s_results} we discuss the implications from our results 
on the understanding of \lya\ emission in LBGs, and on the properties 
of these galaxies.
Our main conclusion are summarised in  Sect.\ \ref{s_conclude}.

\section{Observational constraints from cB58}
\label{s_obs}
The main properties of cB58 of relevance for the present paper
are summarised in what follows.

\subsection{Redshift and ISM outflow properties}   
\label{s_ism}
Direct observations of
photospheric lines by \citet{Pett00,Pett02} assign the systemic redshift
at $z_{\rm UV}=2.7276^{+0.0001}_{-0.0003}$. 
\citet{Tepl04} determine $z_{\rm HII}=2.729\pm0.001$ from rest-frame optical emission
lines(H$\alpha$, [N~{\sc ii}], [O~{\sc iii}] and H$\beta$)
of the counterarc ``A2'', in agreement with their earlier determination
from the main image of cB58.
The CO emission line determination by \citet{Bake04} is $z _{\rm
  CO}=2.7265^{+0.0004}_{-0.0005}$.

The interstellar absorption lines and \lya\ absorption are blueshifted
by $\sim 250$ \kms\ with respect to the galaxy redshift, determined
by its photospheric lines. The small \lya\ emission peak on top of the absorption
(see e.g.\ Fig.\ \ref{cB58_1}) is redshifted  by $\sim 300$ \kms \citep{Pett02}.
Both are evidence for large-scale outflowing medium around the star-forming galaxy. 

Interestingly, and in contrast to the properties seen in the average
LBG spectra of \citet{Shapley03} and spectra of individual $z \sim 3$ 
LBGs observed by \citet{Tapken07}, \lya\ is redshifted in cB58 by a similar amount
as the interstellar lines are blueshifted, instead of being redshifted by 
$\sim + 2 \times \vexp$ with respect to the stellar lines (or by $\sim 3 
\times \vexp$ from the IS lines).
As discussed by \citet{Verh06} the latter behaviour is naturally explained
by a spherically symmetric expanding shell with constant velocity, where 
the emerging \lya\ photons mostly originate from scattering on the backside 
of the shell where they gain redshift of $\sim 2 \times \vexp$.
The smaller redshift of \lya\ indicates already empirically a deviation
from this simple symmetry. This will be substantiated below.
How typical/frequent such velocity shifts are cannot be asserted
from the available data, representing mostly stacked spectra. This will
require many high quality spectra of individual LBGs. However, only the
\lya\ shape, not line shifts, is the criterion used to classify 
cB58 in the quartile with \lya\ absorption.

The low ionisation ISM absorption lines yield consistently a velocity
dispersion $b=70$ \kms \citep{Pett02}. The neutral hydrogen column density of
the outflow is determined by Voigt fitting of the \lya\ absorption
wings;  $\nh = (7.0 \pm 1.5) \times 10^{20}$ cm$^{-2}$ and 
$\nh = 6^{+1.4}_{-2.}\times 10^{20}$ cm$^{-2}$ 
is obtained by \citet{Pett02} and \citet{Sava02} respectively.
Absorption profile line fitting shows that at least $\sim$ 2/3 of the low 
ionisation material is found at $V=-265$ \kms, with a $b$ parameter of 70 \kms\
\citep{Pett02}.

The outflow of cB58 is located well in front of the stars and covers them almost 
completely, since it absorbs almost all the UV light from the background
stars. Indeed, \citet{Sava02} find only a small residual mean flux above the
zero level in the core of the \lya\ absorption feature, whereas it is
black for \citet{Pett02}. \citet{Heck01} estimate an areal covering factor 
for optically thick gas of 98\% from the residual intensity at the core of the 
C~{\sc ii} $\lambda$1335 line.

\subsection{Age and star-formation history of cB58}
UV spectral fitting, including detailed stellar line profile fits, consistently
indicate ages of  $\ga$ 20--100 Myr and favour constant star formation
over this timescale \citep{Elli96,Pett00,deMell00}.
Such an age is also consistent with the observed stellar mass and 
star formation rate \citep{Bake04}.
Another independent argument in
favour of a young age is the underabundance of N and the Fe-peak
elements in the ISM of cB58 \citep{Pett02}, giving the upper limit
age $\la 300$ Myr, the timescale for the release of N from
intermediate-mass stars.  

\subsection{Extinction/dust content}
The precise amount of extinction in cB58 remains unclear. While
\citet{Pett00} estimated $E_s(B-V)=0.29$ from their UV spectral fitting, 
and \cite{Tepl00} calculated $E_g(B-V)=0.27$ from the $\rm
H_{\alpha}/H_{\beta}$ ratio, \citet{Bake01} used four methods
including a correction of the value derived from the Balmer decrement 
obtaining values from $E_s(B-V)=0.042$ to 0.236.
Here $E_s$ and $E_g$ refer to the color excess of the stellar light
and the nebular gas respectively, which are related by $E_s(B-V)=0.44 E_g(B-V)$
if the Calzetti law \citep{Calz00} applies.
From this we conclude that $E_g(B-V) \sim$ 0.10--0.54, and we 
adopt a mean value of $E_g(B-V) \sim 0.3$ for simplicity.

\subsection{Star formation rate, mass, and metallicity of cB58}
The star formation rate estimated from the UV luminosity $L_{1500}$ 
is SFR$_{\rm UV} \sim 40$ \msunyr\ \citep{Pett00} after correcting for
reddening and gravitational magnification (assuming a magnification factor 
$\mu=30$, cf.\ below).
Adopting a lower extinction, based on the Balmer decrement, \citet{Bake04}
derive $SFR_{H_{\alpha}/H_{\beta}}=23.9\pm 0.7 \msunyr$
from \ha.

The velocity dispersion measured from nebular and CO
emission lines are similar: $\sigma_{\rm HII} = 81$ \kms\
\citep{Tepl00}, and $\sigma_{\rm CO} = 74 \pm 18$ \kms\
\citep{Bake04}. The dynamical mass deduced from these values, given
the physical size of 2.4 kpc of the galaxy \citep{Seitz98},
is $M_{\rm dyn} \sim 1. \times 10^{10}$ \msun. 
The mass of the gas deduced from the CO column density is $M_{\rm gas} = (6.6
\pm^{5.8}_{4.3})\times 10^{9}$ \msun \citep{Bake04}.

The metallicity derived from the interstellar absorption lines and
from the ionised region are compatible: $Z_{\rm IS} \sim 2/5$ \zsun\
\citep{Pett02} and $Z_{\rm HII} \sim 1/3$ \zsun\ \citep{Tepl00}.

\subsection{Gravitational lensing}
Lensing reconstruction shows that the primary image of
cB58, an arc whose spectrum will be used here, is magnified by 3.35--4 mag overall, 
i.e.\ a magnification factor $\mu \sim$ 22--40  \citep{Seitz98}.
Approximately only a fraction 0.5--0.66 of the source is imaged 
into the cB58 arc according to \citet{Seitz98,Bake04}. However, the fact that
cB58 follows all the same scaling relations (among \lya\ equivalent width,
interstellar absorption line strengths, extinction etc. \citet{Shapley03})
obeyed by unlensed LBGs, shows that the spectrum is not an unrepresentative
slice of the galaxy. 

\section{Fitting the \lya\ spectrum of cB58}
\label{s_fit}

\subsection{Observations}
We use the normalised rest-frame UV spectrum from \citet{Pett02},
obtained with the Echelle Spectrograph and Imager (ESI) on the Keck 
telescope II at a resolution of 58 \kms, and kindly made available
to us by Max Pettini.
We have also examined the UVES spectrum taken with the VLT from 
\citet{Sava02}. Except for slight differences in the normalisation
the spectra show a very good agreement (cf.\ Fig.\ \ref{fig_synth2}).
The former spectrum is used subsequently for our line profile
fitting.

\subsection{\lya\ plus UV continuum modeling}
To fit the observations we use our 3D Monte Carlo (MC) radiation transfer 
code {\em MCLya} \citep{Verh06}.
The code solves the transfer of \lya\ line and adjacent continuum photons 
including the detailed processes of \lya\ line scattering, dust scattering
and dust absorption.

For simplicity, and given empirical evidence in favour of a fairly
simple geometry in $z \sim 3$ Lyman Break Galaxies (hereafter LBG)
discussed by \citet{Verh06}, we first adopted a simple ``super-bubble'' model
to attempt fitting the observed \lya\ line profile.
The assumed ``standard'' geometry is that of an expanding, spherical,
homogeneous, and isothermal shell of neutral hydrogen surrounding
a central starburst emitting a UV continuum plus \lya\
recombination line radiation from its associated \hii\ region.
The homogeneity and  a covering factor near unity are also supported
by the observations (cf.\ Sect.\ \ref{s_obs}).
Taking into account deviations from a constant expansion velocity 
is necessary to reproduce the observed \lya\ profile of cB58,
as already indicated from the empirical evidence discussed earlier.

\begin{table}[tb]
    \caption{Input parameters of the ``standard'' cB58 model for the radiation 
transfer code}
    \label{tab_pars}
\begin{tabular}{lrllll}
\hline
\hline
\vexp $=V_{\rm front}$ & 255 \kms \\
\vback\ &  free \\
\nh   & $7.0 \times 10^{20}$ cm$^{-2}$ \\
$b$   & 70 \kms \\
E$_g$(B-V) & 0.3 \\
FWHM(\lya) & 80 \kms \\
EW(\lya) & free \\
\hline
\end{tabular}

  \end{table}
\subsubsection{Radiation transfer modeling}
The following geometries have been adopted:
{\em 1)} a spherically symmetric expanding shell with velocity \vexp, 
{\em 2)} a foreground slab moving towards the observer with velocity 
$V_{\rm front}=\vexp$, and
{\em 3)} two parallel slabs with different velocities \vback\ and
$V_{\rm front}=\vexp$ respectively to mimic a shell like geometry allowing for
velocity variations between the front and the back of the shell.
The value of \vexp\ is taken from the observations.
\vback\ is a free parameter.

The other model parameters are: the velocity dispersion given by the Doppler
parameter $b$, the \hi\ column density \nh, and the dust absorption
optical depth $\tau_a$ at \lya\ wavelength. 
We assume that dust and \hi\ are uniformly mixed.
As discussed in \citet{Verh06}, $\tau_a$ relates to the usual
extinction $E(B-V) \approx (0.06 \ldots 0.11) \, \tau_{a}$,
where the numerical coefficient covers attenuation/extinction laws
of \citet{Calz00}, \cite{Seat79} and similar.
Here we assume  $E(B-V) = 0.1 \tau_a$. 
The values for these parameters, summarised in Table \ref{tab_pars}, 
are taken from the observations of cB58 (Sect.\ \ref{s_obs}). 
Note that we suppose that the color excess measured for the nebular
gas is representative for the dust optical depth seen by the \lya\
photons.

The spectrum of the central point source is
described by the stellar continuum around \lya, either a flat
continuum or synthetic starburst spectra described below, and the
nebular recombination line, described by its equivalent width
$EW(\lya)$ and full width at half maximum $FWHM(\lya)$.
Although in principle the equivalent width of the intrinsic 
\lya\ emission is determined by the stellar population, i.e.\ can
be predicted by the synthetic spectra, $EW(\lya)$ is kept as a free
input parameter for the radiation transfer modeling.
The resulting fit values will subsequently be compared to values
expected from synthesis models to examine if consistent solutions
can be obtained.
The value for FWHM is taken from the observed \ha\ line ($\sigma_{\rm HII}$).

For the spherical shell (1), the emergent spectrum is computed by integration
over the entire volume. This assumes that the shell is unresolved and
completely integrated within the spectroscopic aperture, which corresponds
to $\sim$ 2 kpc \citep{Seitz98}.
For the slabs (2), the emergent spectra are computed taking the photons 
emitted within an opening angle of typically 60 degrees into account.
The results depend little on the exact opening angle.

All models have been computed using the Monte Carlo (MC) radiation transfer 
code {\em MCLya} developed by \citet{Verh06}.
The code treats the scattering of line and continuum photons
in the \lya\ line and the adjacent continuum, as well as scattering
and absorption by dust.
In practice one high statistics MC simulation is run for each set of
parameters listed above assuming a constant (flat) input spectrum.  By
keeping trace of the relation between the input frequency bin of the
photons and their emergent frequency, simulations for arbitrary input
spectra can be generated a posteriori in an interactive manner.
This avoids unnecessary transfer computations without resorting to
any simplifying assumption.
For comparison with the observations, the theoretical spectrum is finally
convolved with a gaussian to the instrumental resolution.

Simulations for multi-slab geometries (3) have been computed 
by combining the results from a single slab MC computations
in a numerical fashion, modifying appropriately the relative
velocity shifts of the slabs and opening angles of the emerging
radiation. This allows fast, interactive fitting, without resorting
e.g.\ to approximate analytic fits \citep[cf.][]{Hansen06}.


\begin{figure}[htb]
\includegraphics[width=8.8cm]{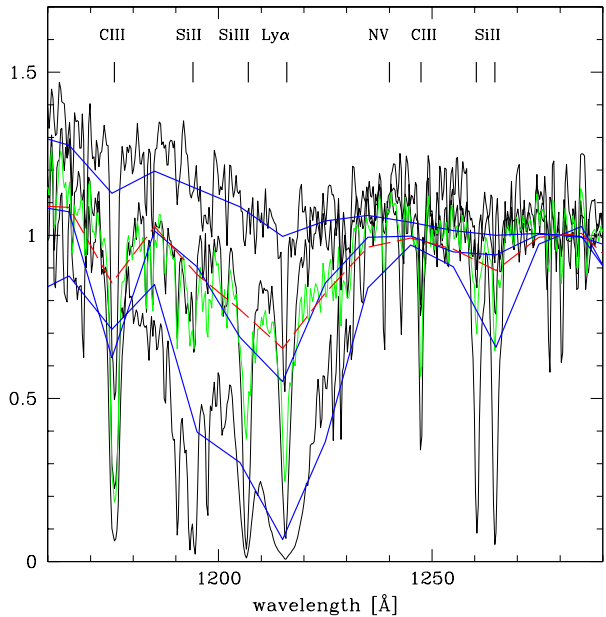}
\caption{Predicted stellar spectra around \lya\ from the synthesis
  models of S03 (blue and red low-resolution curves), and the
  high-resolution models of \protect\citep{Gonz05} in black and green.
Computations for solar metallicity bursts of ages 0.1, 10, and 50 Myr
  are shown from top to bottom. The green line shows the
  equilibrium spectrum reached for constant star formation, as the red
  dashed line for the S03 model. All spectra are in $F_\lambda$ units,
  normalised arbitrarily at 1280 \AA. The main photospheric line
  identifications are indicated at the top (cf.\ text). Note the
  excellent overall agreement of the two independent models, allowing
  a reasonable description of the broad absorption around \lya.} 
\label{fig_synth1}
\end{figure}

\subsubsection{Synthetic starburst spectra in the \lya\ region}
\label{s_synth}
When spatially integrated galaxy spectra are considered, it is 
a priori necessary to include both stellar and nebular lines
in fits of observed spectra. As the region around \lya\
is generally not considered and not included or discussed in
most evolutionary synthesis codes, we here briefly examine
the behaviour of this spectral range.
To do so we use the evolutionary synthesis models of 
\citet{Scha03} (hereafter S03) complemented by the recent models 
of \citet{Gonz05} including high spectral resolution stellar atmospheres,
kindly made available to us by Miguel Cervi\~no.
Only the main ingredients of these model sets shall be summarised here.

The S03 models basically include simplified non-LTE model atmospheres
for massive stars ($>20$ \msun), and LTE line blanketed Kurucz
atmospheres otherwise.  The \citet{Gonz05} models include
in particular theoretical high spectral resolution libraries, and
all massive stars are treated with the plane parallel non-LTE
blanketed TLUSTY code \citep{Lanz03}.  In both cases the
same stellar tracks and metallicities, summarised in S03, have been
used. With respect to the \lya\ region these models represent
the most up-to-date computations. The only restriction is that
these computations do not treat the stellar winds of the hottest stars,
which leads to some modifications of the intrinsic \lya\
line profile of these stars. However, unpublished test computations
\citep{Mart00} show that these effects are small or negligible
compared to the underlying \lya\ absorption from the bulk of B and A stars
and compared to the expected nebular emission.
This is also confirmed by \citet{Leitherer07}.

\begin{figure}[htb]
\includegraphics[width=8.8cm]{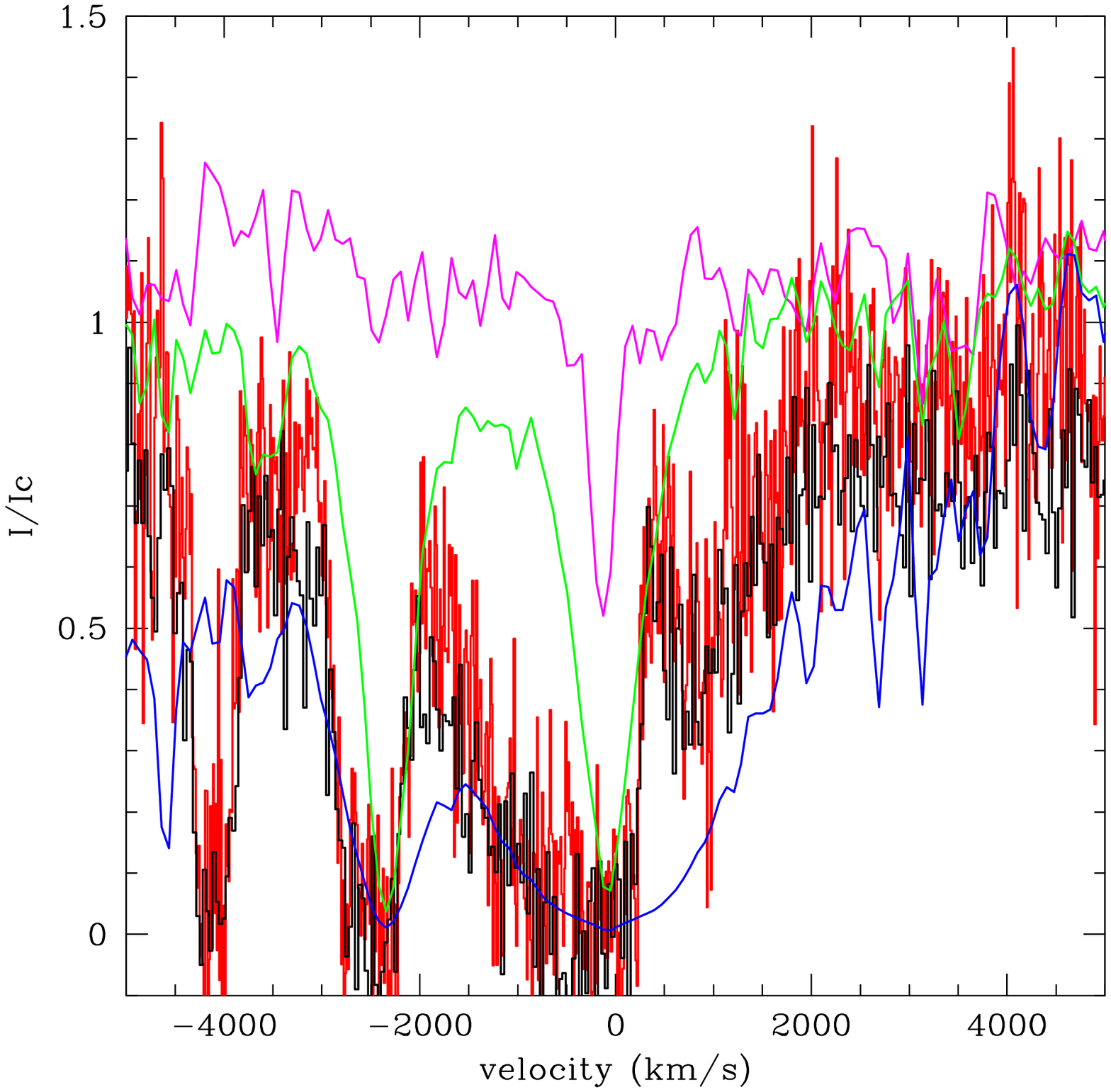}
\caption{Observed spectra of cB58 from \protect\citet{Pett02} (black)
and \citet{Sava02} (red) compared to the high-resolution synthesis models
of \protect\citep{Gonz05} for solar metallicity bursts of ages 0.1, 10, and 50 Myr
(in magenta, green and blue from top to bottom).
Apart from small differences in the normalisation the two observed spectra 
are very consistent. The comparison with the models illustrates how the
shape of the red wing of \lya\ can be clearly used to exclude bursts of ages
larger than 10--50 Myr.}
\label{fig_synth2}
\end{figure}

\begin{figure}[htb]
\includegraphics[width=8.8cm]{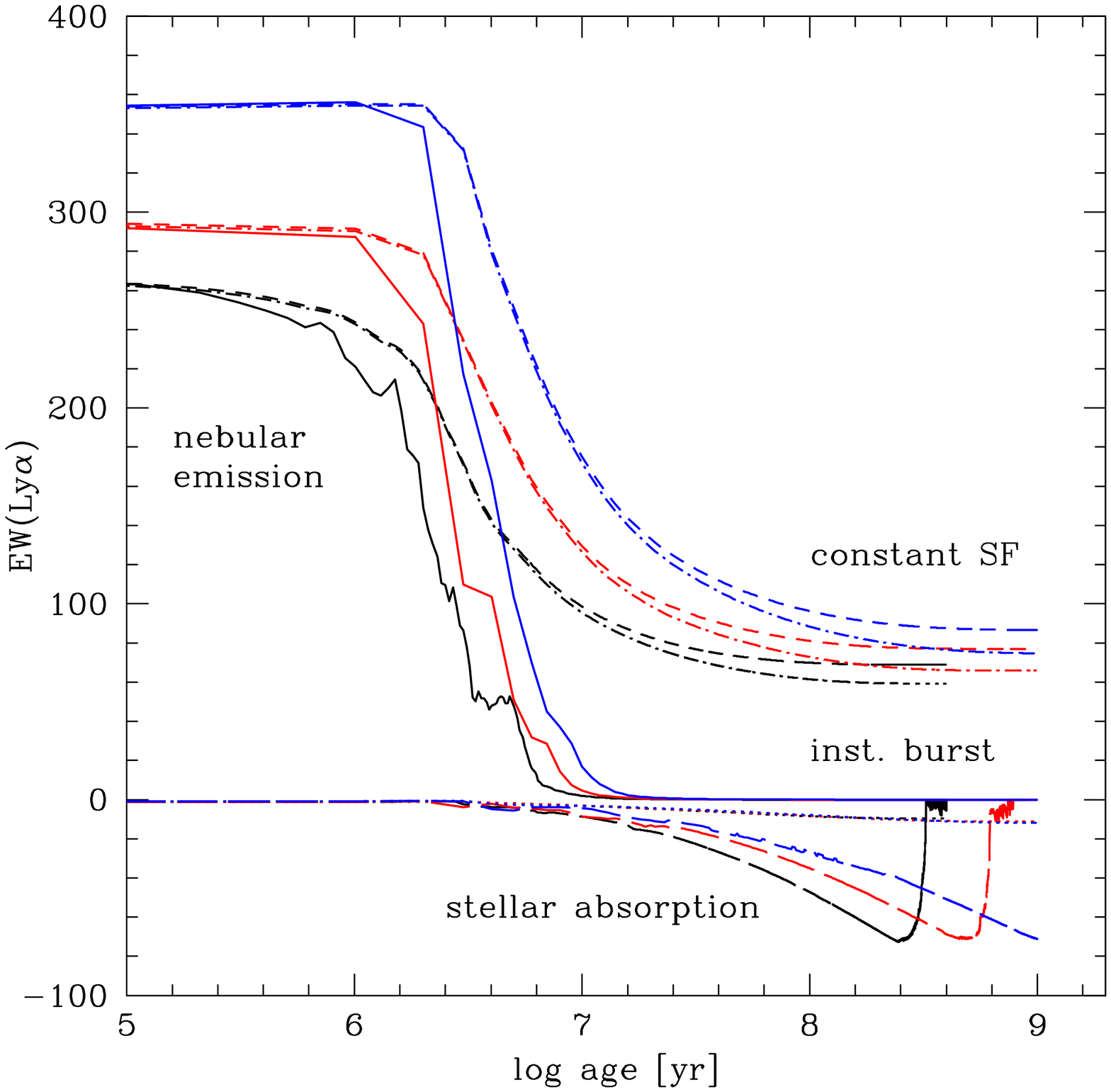}
\caption{Predicted \lya\ equivalent widths of the nebular emission ($EW(\lya)>0$)
and stellar absorption ($EW(\lya)<0$) for instantaneous bursts (solid and long-dashed 
lines) and  models with constant star formation (short dashed, dotted, and dash-dotted).
The long-dashed lines show the total $EW$ (nebular $+$ stellar) for constant
star formation, the dotted lines the stellar absorption component.
Three different metallicities, $Z=0.02$ (solar, black), 0.004 (red), and 0.0004 (blue)
are shown.}
\label{fig_ewlya}
\end{figure}

In Fig.\ \ref{fig_synth1} we show the spectra predicted by these two
synthesis over the spectral range centered on \lya.
The spectra, computed for instantaneous bursts with ages 0.1, 10, and 50 Myr
at solar metallicity and for a Salpeter IMF with an upper mass cut-off
of 100 \msun, are normalised arbitrarily at 1280 \AA.
The high-resolution spectra allow the identification of the main
(photospheric) spectral lines, which are
the C~{\sc iii} multiplet centered at 1175.6 \AA, 
Si~{\sc ii} multiplets at $\lambda\lambda$ 1193.3,1194.5 and
1260.4,1264.7, the Si~{\sc iii} multiplet  $\lambda\lambda$ 1206.5,1207.5,
and C~{\sc iii} $\lambda$1247.4. The position of the N~{\sc v} multiplet
$\lambda\lambda$ 1238.8,1242.8 formed in the winds of early type stars 
(not included in the present models) is also indicated.

The temporal behaviour of the \lya\ profile in a burst can be sketched
as follows: 
for ages $\la$ 10 Myr the \lya\ absorption deepens but remains relatively
narrow. At older ages, when stars with a lower ionisation degree in their
atmospheres dominate, the \lya\ becomes considerably stronger and broader.
At $> 50$ Myr, not shown here, the red wing remains broad, whereas the flux
shortward of \lya\ disappears very rapidly.
For constant (ongoing) star formation the spectrum reaches quite shortly
(over $\ga$ 10 Myr) an equilibrium with a \lya\ line profile quite closely
resembling that of a $\sim$ 10 Myr burst, as also shown in Fig.\
\ref{fig_synth1}. 

The shape of the red wing of \lya, up to $\sim$
1240 \AA, provides a useful constraint on the age and star formation history
of the starburst. For bursts with ages $>>$ 10 Myr the width of the stellar \lya\
absorption become so large, that a measurement of the shape of its
red wing can already be distinctive, even without considering interstellar
\lya\ radiation transfer. For example, for the case of cB58
shown in Fig.\ \ref{fig_synth2} the observed \lya\ wing is clearly
less broad than the burst model at 50 Myr. Younger bursts or a more constant
star formation history are therefore required.
Of course, other UV spectral lines such as the well known Si~{\sc iv} and 
C~{\sc iv} lines, are sensitive to age and star formation history
\citep[cf.][]{Leitherer95}.

As other parts of the UV spectrum \citep[see e.g.][]{SB99}, the
detailed integrated spectrum in the \lya\ region depends also somewhat
on metallicity (cf.\ Fig.\ \ref{fig_ewlya}). The main reason is the
systematic shift of the average stellar effective temperature with
metallicity for a given age. In consequence similar spectra of bursts
can be obtained at different metallicities, provided some age shift is
allowed for.  Despite the detailed metallicity differences, which can
be accounted for by using the appropriate high resolution spectra, the
overall behaviour of the shape of the \lya\ profile remains quite
independently of metallicity, as could be expected.

It is also reassuring to note the good overall agreement of the two
independent synthesis models, allowing a reasonable description of the broad
absorption around \lya\ even with low-resolution spectra. In any case
predictions from both synthesis models, as well as flat spectra, will
be used below as input for the cB58 fits with the \lya\ radiation
transfer code.

The relative strengths of the stellar \lya\ absorption and
expected intrinsic nebular emission is illustrated in Fig.\
\ref{fig_ewlya}, where the predicted \lya\ equivalent widths of these
components are plotted as a function of time for the case of
instantaneous bursts and for constant star formation, as well as for
three different metallicities. The behaviour of these quantities is as
expected from earlier computations at solar metallicity
\citep{Charlot93,Valls93}.  Decreasing metallicity increases the
nebular emission and introduces a small age shift in \lya\ absorption.
For constant star formation over $\sim$ 20--100 Myr or more, one
expects in particular $EW(\lya) \sim$ 60--100 \AA.

\section{Results}
\label{s_results}
We now discuss the simulated spectra assuming different geometries
and their dependence on the input parameters.

\begin{figure}[tb]
\includegraphics[width=8.8cm]{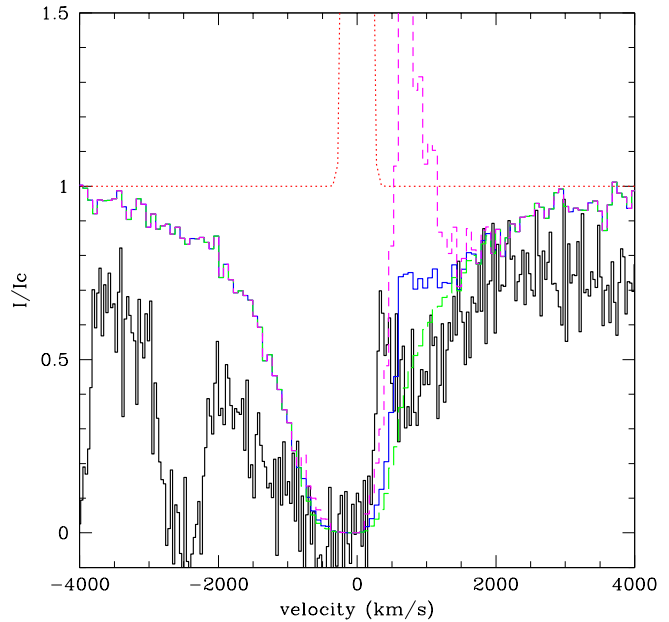}
\caption{Predicted \lya\ line profile for a symmetric spherical shell with
constant velocity, the input parameters summarised in Table \ref{tab_pars},
and EW(\lya)=0, 10, 40 \AA\ (green, blue, magenta respectively). 
The input spectrum is the red dotted line. 
The observed spectrum of cB58 from \citet{Pett02} is shown in 
black.}
\label{cB58_1}
\end{figure}

\begin{figure}[tb]
\includegraphics[width=8.8cm]{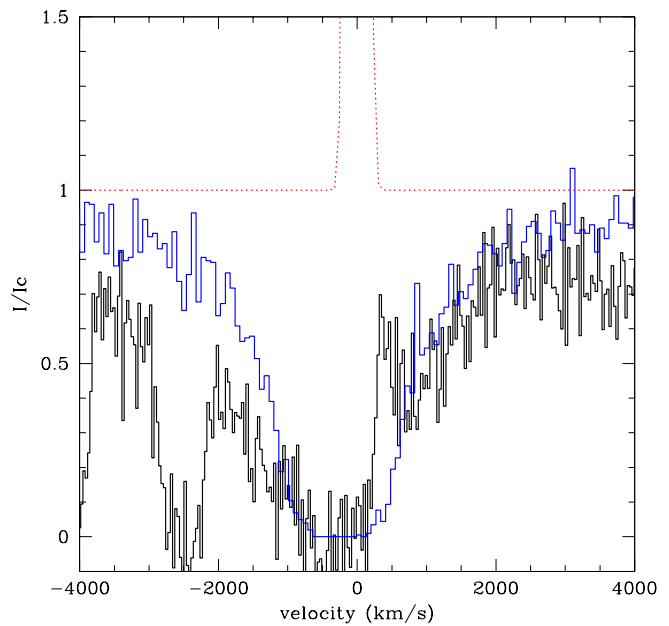}
\caption{Observed (black) and predicted \lya\ line profile for the 
single foreground slab model (blue). The input spectrum, shown by
the red dotted line has EW(\lya)=40 \AA.}
\label{cB58_1slab}
\end{figure}
\subsection{Models for a constant velocity shell}
The predicted \lya\ line profile for a symmetric spherical shell with
constant velocity, computed with the input parameters summarised
in Table \ref{tab_pars} and with EW(\lya)=0, 10, and 40 \AA\ is 
compared to the observed profile in Fig.\ \ref{cB58_1}. 
The input spectrum, here a flat continuum plus line, is also shown 
for illustration.

As expected from our earlier simulations \citep{Verh06}, 
the global shape of the theoretic spectrum is similar 
to the observed one, consisting of a broad damped absorption
plus a redshifted asymmetric emission line component, whose strength
depends on the strength (EW) of the input line.
The net absorption is due to dust absorption, since the case
considered here is the integrated spectrum of a shell, where 
all scattered photons are observed. Scattering of photons out of
the line of sight is therefore not possible, hence any net absorption
must be due to dust. 
The superposed emission peak results from the radiation transfer of the
photons from the original (input) emission line, which are scattered and 
absorbed, and are able to escape only thanks to their redshift acquired 
on the back of the shell.

Two main problems are immediately clear from Fig.\ \ref{cB58_1}.
First the predicted width of the absorption is unsufficient, and second
the position of the predicted \lya\ emission peak is found at too high positive 
velocities.
The first problem can be solved by increasing the column density
of the shell. As already mentioned above, and first pointed out by
\citet{Verh06}, an integrated shell geometry necessarily requires a
larger \nh\ to reproduce a simple Voigt profile of the same width
as a slab of the same column density.
The second discrepancy with the simple shell model could also be anticipated
(cf.\ Sect.\ \ref{s_obs}), since the redshifted peak of \lya\ is expected
at $2 \times \vexp$, whereas the observations show the peak around
$\sim 300$ \kms.
As the velocity shift of the \lya\ peak is a fairly generic feature of 
symmetric expanding shell models, it is clear that this discrepancy cannot
be resolved by changes of the input parameters, such as \nh, $b$, E(B-V),
etc. This has been confirmed by numerous model calculations.

Note also that this model allows only for a fairly weak intrinsic
\lya\ emission with EW(\lya) $<$ 10 \AA, much lower than the 
value expected from synthesis models for the star formation history
derived for cB58 (cf.\ below). 

\subsection{Single slab model}
\label{s_1slab}

Given the difficulties with a simple shell model and the fact that
the direct observational constraints on the outflow only relate to
cold material in front of the UV source, one could ask the question whether
a single moving foreground slab (without any backscattering source)
could explain the observations.
In other words this boils down to the question if the observed velocity
shift of $\sim$ 255 \kms\ between stellar and interstellar absorption
lines could be enough to allow for a partial transmission of the 
intrinsic \lya\ line, such that it appears at velocities $\ga 250$ \kms?
The simple answer obtained from many simulations is no. More precisely,
with the observed column density, $b$ parameter, E(B-V), and FWHM(\lya) 
the transmission of \lya\ line photons emitted at line center ($V=0$) through a slab
is negligible, and hence no line is formed on top of the broad
absorption line profile, independently of the intrinsic \lya\
equivalent width. 
This is shown in Fig.\ \ref{cB58_1slab}.

For enough line photons to escape trough the slab
the intrinsic emission line must be redshifted by $> 200$ \kms, which is
larger than the possible redshift of $\sim$ 100 \kms\ found by \citet{Tepl00}
from 10 restframe optical nebular emission lines\footnote{As pointed 
out by \citet{Pett02} this possible small reshift could well be 
due to uncertainties in the absolute wavelength calibration.}.
In any case, such a solution would require a large intrinsic \lya\ equivalent 
width (EW(\lya) $\ga$ 40 \AA) in agreement with the results
found below.

Hence the presence and position of the \lya\ peak together with a relatively
large HI column density testified by the damped absorption 
excludes most likely a pure foreground slab. To allow for the \lya\
transmission, a receding background screen must be present so that 
the \lya\ photons can gain an additional redshift by bouncing off the back,
in a similar fashion as in the spherical shell. Such a situation
will be simulated now.

\begin{figure}[tb]
\includegraphics[width=8.8cm]{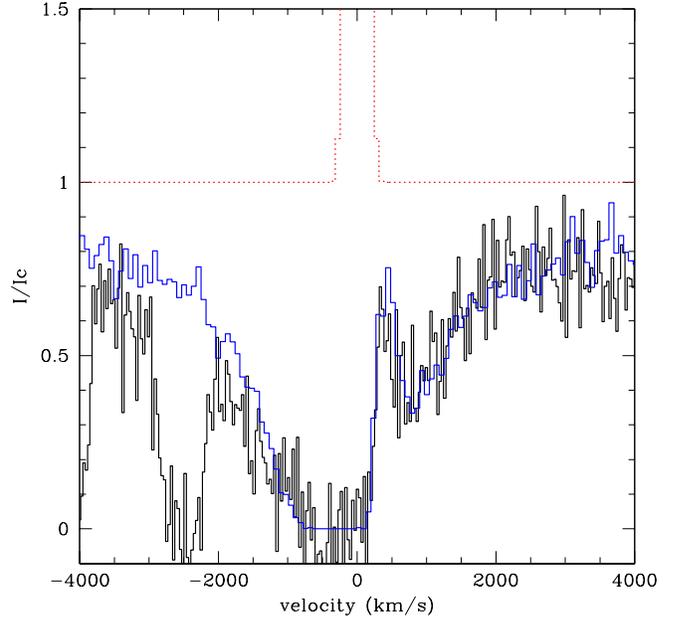}
\caption{Observed (black) and predicted \lya\ line profile for the 
two slab model described in Sect.\ \ref{s_2slabs} (blue). The input spectrum
is shown by the red dotted line.}
\label{cB58_2slabs}
\end{figure}

\begin{figure}[tb]
\includegraphics[width=8.8cm]{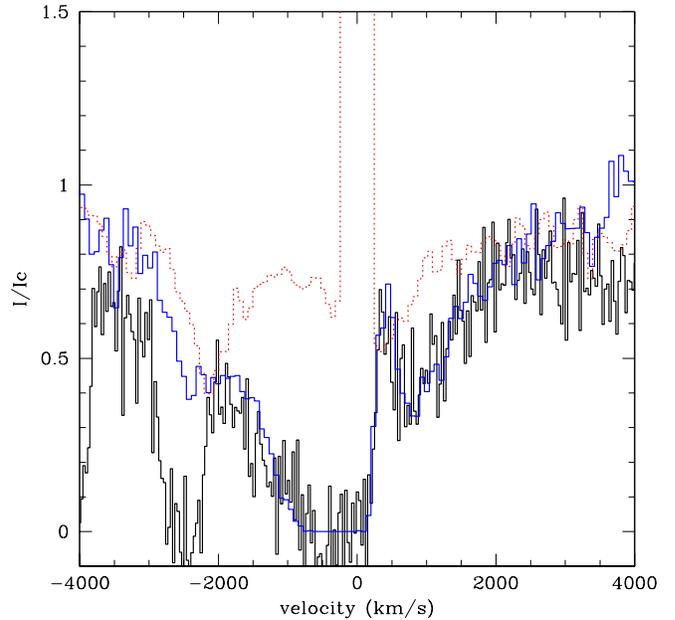}
\caption{Same as Fig.\ \protect\ref{cB58_2slabs} but using 
high resolution spectra from evolutionary synthesis models
for the description of the stellar continuum.  The input spectrum,
arbitrarily normalised, is shown by the red dotted line.}
\label{cB58_2slabs_cervino}
\end{figure}

\subsection{Models for two moving slabs}
\label{s_2slabs}
To attempt reproducing the observed \lya\ emission component we now
assume a geometry with two parallel slabs, one in front and one in the back
of the UV and \lya\ source. While the velocity and the other parameters
of the approaching foreground slab are fixed by the observations (cf.\ Table
\ref{tab_pars}), the velocity \vback\ of the receding slab is kept as 
a free parameters.

A good fit to the observed spectrum of cB58 is found for
\vback\ $\sim$ 140 \kms, as shown in Fig.\ \ref{cB58_2slabs}.
For this choice of \vback\ the velocity shift of the receding
slab is sufficient to allow a non-zero transmission of the reflected
\lya\ photons through the foreground slab, and low enough to
produce the peak at the observed velocity of $\sim 300$ \kms.
The emergent spectrum is basically independent on the properties
(e.g.\ \nh) of the background slab, which merely acts as a moving mirror.
This is clear since the spectrum reflected from a slab depends
only on the thermal width $b$, and only weakly so ($\propto b^{-1/2}$),
as e.g.\ shown by \citet{Hansen06}.

Interestingly the value of the second free parameter, the intrinsic \lya\ equivalent
width EW(\lya), is required to be EW $\ga$ 80 \AA\ to reproduce the 
observed strength of the \lya\ emission peak.
In our two slab model the solid angle under which the receding ``mirror''
is seen by the foreground slab governs the required EW, since
this controls the relative amount of photons received from direct
and reflected radiation at the back of the foreground slab.
The lower limit on EW(\lya) given above is obtained by assuming
a solid angle of $2 \pi$.
The observations are compatible with arbitrarily larger values of EW(\lya).

Instead of the flat continuum we have also examined models using the
more realistic synthetic spectra including stellar \lya\ absorption
presented above (Sect.\ \ref{s_synth}) plus nebular \lya\ emission as
input for the radiation transfer calculations. The resulting \lya\
profile, shown in Fig.\ \ref{cB58_2slabs_cervino}, is of similar quality
for a slightly lower input \lya\ equivalent width of $\ga$ 60 \AA.

Fits using different combinations of the column density and extinction
could be examined, given the uncertainty on the dust optical depth
discussed above  (cf.\ Sect.\ \ref{s_obs}).
Using e.g.\ lower values of $E_g(B-V)$ and our ``standard'' value
for \nh\ leads to a narrower \lya\ absorption profile, which would
require a higher column density. Quantifying the degeneracies
around the solution found here is beyond the scope of this paper.
In any case, it is quite clear that the main conclusion, a high column
density and the need for a large intrinsic \lya\ equivalent width, 
must be quite robust with respect to these degeneracies.
This must be the case since a high column density is required to 
create the broad \lya\ absorption (independently of the exact geometry), 
and in this case the escape fraction of \lya\ photons ``injected'' close 
to line center is low quite independently of the exact dust content.

In summary, we conclude that with the simple geometry considered here
we are able to reproduce for the first time the observed \lya\ profile
of cB58 with a radiation transfer model using all observed constraints
on the outflow plus just two free parameters, \vback\ and EW(\lya).
One of the interesting conclusions from this analysis is to show 
that the observed absorption dominated \lya\ line profile of cB58
is compatible with an intrinsically very different spectrum with 
a strong (EW(\lya) $\ga$ 60 \AA) \lya\ emission line.
Radiation transfer effects and the presence of dust transform
the intrinsic starburst spectrum to the observed one.

\section{Discussion: a unifying scenario for \lya\ emission in LBGs}
\label{s_discuss}

As we have shown above the observed \lya\ spectrum of cB58 is compatible
with an {\em intrinsically} strong \lya\ emission (with EW(\lya) $>$ 60 \AA).
Hence this is compatible with the equivalent width expected for a starburst with a
constant star formation rate over $>$ 50--100 Myr, the timescale over which
EW(\lya) reaches a constant equilibrium value of $\sim$ 60--80 \AA\ for
a ``standard'' IMF and metallicities comparable to that of cB58 \citep[see Fig.\
\protect\ref{fig_ewlya} and][]{Scha03}.
Therefore the observed \lya\ strength and profile are compatible with 
the star-formation histories derived from detailed modeling of the rest-UV 
properties, who consistently found a constant star formation rate and ages up to
ages of $\sim$ 20--100 Myr \citep[cf.][]{Elli96,Pett00,deMell00}.
Constant star formation over such timescales, instead of an aged burst which 
would be responsible for a low EW(\lya), is also physically more plausible
for objects like cB58 having large star formation rates, large physical scales,
etc. Taking into account \lya\ radiation transfer and dust allows us to reconcile
these diagnostics.

Both detailed analysis of the rest-UV spectra of some LBGs and broad band 
SED fitting of many LBGs yields typically stellar ages of several hundred
Myr and fairly constant or slowly decreasing star formation rates
\citep[cf.][]{Shapley01,Papovich01,Pentericci07}.
These results and our analysis of cB58 indicate fairly long star formation
timescales, in contrast e.g.\ to the suggestion of \citet{Ferrara06}.

cB58 is part of the quartile of $z \sim 3$ LBGs showing 
a \lya\ dominated by absorption with faint or little \lya\ emission 
\citep{Shapley03}.
Therefore we can expect that these objects behave in a similar way as
cB58 and hence have much higher {\em intrinsic} \lya\ equivalent
widths and higher {\em intrinsic} \lya\ luminosities. In other words
observed distribution functions of EW(\lya) and the
\lya\ luminosity function must be strongly modified by radiation
transfer and dust effects, and the fraction of objects with low EW(\lya)
($L(\lya)$) must be ``artificially'' overestimated.

Taking radiation transfer and dust into account, should therefore
allow to reconcile observed \lya\ strengths with relatively long
($\gg 50$ Myr)  ``duty cycles'' of LBGs.
In any case, age differences between $\sim$ 200 and 400 Myr
found by \citet{Pentericci07}  for LBGs with and without \lya\ emission 
respectively, cannot be the direct cause of the observed \lya\ 
differences.
If our scenario also works for \lya\ emitters (henceforth LAEs, i.e.\ 
\lya\ selected LBGs) and the suppresion of \lya\ photons is sufficient,
it could explain the relatively low percentage of star-forming 
galaxies with no/little \lya\ emission and hence reduce the need for short
starburst episodes invoked e.g.\ by \citet{Malhotra02}.
This would then also help to reconcile the lower number densities of LAEs compared
to LBGs despite the agreement in their bias \citep{Kovac07},

It may be reasonable to expect that most of the LBGs have 
quite very similar {\em minimum intrinsic} \lya\ equivalent widths
of $\sim$ 60--80 \AA\ determined by constant star formation over 
$>$ 50--100 Myr. 
Larger EWs are expected for objects dominated by 
younger ($\la$ 10--40 Myr) populations (cf.\ Fig.\ \ref{fig_ewlya}).
Lower EWs would then be expected 
either in post-starburst objects or due to transfer and dust effects.
Since the outflow velocities of LBGs do not vary strongly between
subsamples with different \lya\ strengths\footnote{The expansion 
velocities determined by one third of the velocity shift between
the interstellar absorption lines and the \lya\ peak, 
$\vexp \sim 1/3 \Delta(V_{\rm ISM},V(\lya))$ is $\sim$ 160--260 \kms\
for the sample of \citet{Shapley03}.}
we propose that the main
factor leading to the diversity of \lya\ strengths in LBGs is their 
\hi\ column density and concomitantly their dust content
\footnote{If the dust-to-gas ratio remains constant, both \nh\ and
the dust content increase in lockstep.}.

This suggestion is compatible with the observational correlations
found between E(B-V) and EW(\lya) \citep[e.g.][]{Shapley03,Tapken07} and 
with the differences in extinction found by \citet{Pentericci07} between
two subsamples of $z \sim 4$ LBGs.
It also broadly agrees with \lya\ line profile fitting of
LBGs with strong \lya\ emission, discussed in the next paper
of this series \citep{Verh07}.

Given observed mass-metallicity relations
\citep[e.g.][]{Tremonti04,Erb06} it may be quite natural to speculate
that the difference in \nh\ and dust content is related to the galaxy
mass to first order. The data of \citet{Shapley01} show such a trend.
In this case more massive LBGs would be more dusty, and hence show
lower EW(\lya), although they could have identical \lya\ emission
properties intrinsically. Such a trend is e.g.\ found by
\citet{Pentericci07}. 
The absence of bright LBGs (in the UV restframe) with high \lya\
equivalent widths (in emission) reported in many papers
\citep[e.g.][]{Shapley03,Ando04,Tapken07} would also naturally be
explained by this fact, if the SFR correlates with mass, as 
observed for star forming galaxies out to $z \sim 2$
\citep[see][]{Elbaz07,Daddi07,Noeske07}.

Relevant for determining the \lya\ properties are however, the total
\hi\ column density and the relative dust to \hi\ content. Assuming the latter
to be constant\footnote{Combining e.g.\ the dependence of the dust-to-gas ratio 
on metallicity O/H from \citet{Lisenfeld98} with the mass-metallicity relation
at $z \ga 2$ \citep{Erb06}, gives a very small dependence of the dust-to-gas ratio 
on galaxian mass.}, the above suggestion would imply larger amounts of \hi\
in more massive LBGs, known to hold at least in nearby galaxies \citep{Kennicutt98}.
However, since the \hi\ responsible for shaping \lya\ is found 
in outflows it may only trace a fraction of the total neutral hydrogen 
content. Hence the \hi\ column density derived from the \lya\ fits
may not be a good tracer of the gas mass, and hence may not simply scale with
galaxy mass. Correlating the outflow properties more closely to those 
of the host galaxy may require dynamical modeling \citep[e.g.][]{Ferrara06,
Blaizot07}.

While the ``scenario'' proposed here to explain the diversity of \lya\
line strengths in LBGs seems consistent with the observations, it remains
to be tested more generally. LBGs with strong \lya\ emission have also
been studied with our radiation transfer models for this purpose and
are broadly in agreement with the scenario proposed here \citep{Verh07}.
Alternative explanations have also been proposed.
For example \citet{Shapley01} have from empirical evidence suggested age as 
the main difference in LBGs, where younger objects are more dusty, and 
hence show less \lya\ emission.
However, it is not clear why older LBGs would contain less dust, especially
since outflows, supposedly used to expel the dust, are ubiquitous
in all LBGs and since precisely these outflows are the location where
the emergent \lya\ spectrum is determined. Also, why would younger LBGs
be more massive than older ones as their data would imply ?
\citet{Ferrara06} have proposed that LBGs host short-lived ($30 \pm 5$ Myr) 
starburst episodes, whose outflows -- when observed at different 
evolutionary phases -- would give rise to the observed correlations 
between IS lines and \lya.
However, their conclusion may need to be revised since the ``observed''
wind velocity versus SFR relation (their Fig.\ 1) is incorrect and 
flatter than assumed\footnote{They assume $v_w=1/2 \Delta(V_{\rm ISM},V(\lya))$
instead of the weaker dependence $v_w \sim 1/3 \Delta(V_{\rm ISM},V(\lya))$ obtained 
from shell models and supported by direct measurements of stellar, IS, and
\lya\ velocity shifts. Furthermore their SFR values differ from those 
quoted in \citet{Shapley03}.}.
Furthermore the ages obtained from SED and spectral fits of LBGs show
older ages and fairly constant star formation histories, as already mentioned
earlier.

In paper III of this series \citep{Verh07} we will examine how
our models are able to explain various correlations observed among
\lya, IS lines, and other properties of LBGs \citep{Shapley03} and
LAEs. Further work will be needed to test the scenario proposed here,
to understand more precisely the relations between LBGs and LAEss,
and to understand the links between star-formation, host galaxy, and
outflow properties.

\section{Conclusions}
\label{s_conclude}
A 3D \lya\ and UV continuum radiation transfer code \citep{Verh06} has
for the first time been applied to the prototypical Lyman break galaxy
MS 1512--cB58 at $z=2.7$. 
Since this is one of, if not the best
studied LBG, and since it is part of the quartile of LBGs
showing predominantly \lya\ in absorption \citep[cf.][]{Shapley03}, 
a detailed \lya\ line profile analysis including radiation transfer and dust
effects is of great interest.

Three different geometries were explored for the material surrounding 
the central starburst; a spherically symmetric
expanding shell with velocity \vexp, a foreground slab moving
towards the observer, and a geometry mimicing a shell with
a lower velocity in the back.
All available observational constraints were used (see Table 
\ref{tab_pars}). In particular these include: the velocity \vexp\ of the 
foreground material, its Doppler parameter $b$, column density \nh,
and the observed extinction.  The width of the intrinsic \lya\ emission
was taken from the observed FWHM of \ha. The two free parameters were:
the equivalent width of the intrinsic \lya\ emission, and the velocity
\vback\ of the background slab (if applicable).

Our radiation transfer calculations have confirmed what could be expected
from our earlier expanding shell simulations and from the lower-than-usual 
observed velocity shift between the peak of ``remnant''
\lya\ emission and the blue IS absorption lines (cf.\ Sect.\ \ref{s_ism}); 
the \lya\ line profile of cB58 cannot be reproduced by an expanding
shell with isotropic constant velocity \vexp. Considering, however, a
lower shell velocity in the back allowed us to obtain excellent fits
to the observed \lya\ line profile.  Interestingly, the input spectrum
requires a fairly strong \lya\ emission, with a lower limit of
EW(\lya) $\ga 60$ \AA. Radiation transfer and the suppresion of
photons by dust are responsible for transforming such an input
spectrum into the observed \lya\ absorption profile with a superposed
faint emission peak. In this way the observed \lya\ line strength and
profile can be reconciled with the strong intrinsic \lya\ emission
expected from the approximately constant star-formation history of
cB58 derived from earlier detailed UV spectral analysis
\citep[cf.][]{Pett00,deMell00}.

In fact we suggest that cB58 and most other LBGs have intrinsically
EW(\lya) $\sim$ 60--80 \AA\ or larger, and that the main physical
parameter responsible for the observed variety of \lya\ strengths and
profiles in LBGs is \nh\ and the accompanying variation of the dust
content (see Sect.\ \ref{s_discuss}).  This explains not only the
absorption-dominated object cB58, but also observed correlations
between E(B-V) and EW(\lya) in LBGs, the absence of bright LBGs with
strong \lya\ emission, and other correlations.

Among the implications from our work are that observed EW(\lya) distributions
and \lya\ luminosity functions must be corrected for radiation transfer
and dust effects. Furthermore relatively short duty cycles, suggested earlier
in the literature, are not required to explain the variations observed
between different LBG types.
Our proposed unifying scenario will be detailed further and subjected
to additional tests in a subsequent publication on a sampe of LBGs 
with strong \lya\ emission \citep{Verh07}. 
Providing a clearer picture of the physical links between the 
observables including \lya\ line strength and profile, and the 
star-formation, host galaxy, and outflow properties and evolution of 
LBGs and LAEs remains an objective for the near future.

\acknowledgements
We thank Max Pettini and Sandra Savaglio for providing us with their 
high resolution UV spectra of cB58, and Miguel Cervi\~no for 
synthetic high-resolution spectra.
This work was supported by the Swiss National Science Foundation. 
\bibliographystyle{astron}
\bibliography{references}

\end{document}